\documentclass[prl,superscriptaddress,showpacs,twocolumn]{revtex4-1}
\usepackage{amssymb}
\usepackage{graphicx}
\usepackage{dcolumn}
\usepackage{bm}
\usepackage{amsmath}
\usepackage{epstopdf}

\begin{document}

\title{Nodal Line Topological Superfluid and Multiply Protected Majorana Fermi Arc in a Three-Dimensional Time-Reversal-Invariant Superfluid Model}
\author{Beibing Huang}
\email{hbb4236@ycit.edu.cn}
\affiliation{Department of Physics, Yancheng Institute of
Technology, Yancheng, 224051, China}
\author{Yujie Bai}
\affiliation{Department of Physics, Yancheng Institute of
Technology, Yancheng, 224051, China}
\author{Ning Xu}
\email{nxu@ycit.cn}
\affiliation{Department of Physics, Yancheng Institute of
Technology, Yancheng, 224051, China}

\date{\today}

\begin{abstract}
{ We theoretically study a time-reversal-invariant three-dimensional
superfluid model by stacking in $z$ direction identical bilayer
models with intralayer spin-orbit coupling and contrary Zeeman
energy splitting for different layer, which has been suggested
recently to realize two-dimensional time-reversal-invariant
topological superfluid. We find that this model shows two kinds of
topologically nontrivial phases: gapless phases with nodal lines in
pairs protected by chiral symmetry and a gapped
phase, both of
which support time-reversal-invariant Majorana Fermi arc (MFA) on
the $yz$ and $xz$ side surface. These MFA abide by time-reversal and
particle-hole symmetries and are topologically protected by the
winding numbers in mirror subspaces and $Z_2$ numbers of two-dimensional DIII class topological
superfluid, different from MFA in the time-reversal broken Weyl
superfluid protected by nonzero Chern number. This important
observation means that MFA in our model represents a new type of
topological state not explored previously. The Zeeman field configuration in our model is relevant to antiferromagnetic topological insulator MnBi$_2$Te$_4$, thus our work stimulates the further studies on superconducting effects in the realistic antiferromagnetic topological insulator.}
\end{abstract}

\maketitle

\vskip 1cm

\section{Introduction} Accompanying the hottest discussion on gapped
topological phase \cite{hasan, Qi, Schnyder08, Kitaev}, gapless
topological phase with Fermi surface has gradually become another
studying focus \cite{horava, volovik}. Considering all ten
Altland-Zirnbauer symmetry classes, topologically stable Fermi
surfaces with all kinds of codimension $p$ have been classified
according to their topological charges from the K-theory
\cite{wangzd1, wangzd2}. A typical representative explored in
various cold atom systems \cite{Gong, zhangchuanwei1,
zhangchuanwei2, vincent, huhui, liuxiaji, ketterler, jianghua} and
solid state materials \cite{wanxiangang, hongming, shiming, suyang1,
bqlv, suyang2, suyang3, nxu, zhenwang, balents, meng, sau,
fangzhong1} is three-dimensional (3D) Weyl semimetal/superfluid with
codimension $p=3$, which characterize Weyl point (WP) Fermi
surfaces. The WP can be regarded as the magnetic monopole in momentum
space \cite{fangzhong2, hosur, cheny} and topologically protected by
nonzero Chern number defined for any two-dimensional (2D) surface
enclosing the WP.

The bulk-boundary correspondence signifies the nontrivial surface
states in Weyl semimetal/superfluid. For any surfaces not
perpendicular to lines connecting WPs, there exists open Fermi arc,
which connects projections of the WPs with opposite Chern number on
the surface Brillouin zone \cite{wanxiangang}. In terms of Weyl
superfluids, Fermi arc possesses Majorana character and is also
denoted by Majorana Fermi arc (MFA) due to inherent particle-hole
symmetry (PHS). To our knowledge in all existing models realizing
the Weyl superfluid \cite{Gong, zhangchuanwei1, zhangchuanwei2,
vincent, huhui, liuxiaji, meng, sau}, time-reversal symmetry (TRS)
is explicitly broken, and MFA is protected by nonzero Chern number.

In this work, we suggest a 3D theoretical model to study
MFA not protected by nonzero Chern number. Our main results are as
followings. The suggested model realizes two kinds of topologically
nontrivial phases, both of which support MFA. The first kind of
phase is gapless. Different from Weyl superfluids, this
gapless phase has nodal lines instead of WPs. The topological
stabilities of nodal lines are protected by chiral symmetry
\cite{wangzd1, beri, sato}. By tuning the parameters of the model,
these nodal lines disappear in pairs, driving the system into a
fully gapped phase, which shows the MFA in the whole Brillouin zone in the $z$ direction and is different from the conventional gapped topological phases. Moreover MFA in two kinds of phases are topologically protected by the winding numbers
in mirror subspaces and $Z_2$ numbers of 2D
DIII class topological superfluid, but not nonzero Chern number. This important
observation means that MFA in our model represents a new type of
topological state not explored previously.

\section{Theoretical model} The model we consider can be regarded as
3D extension of a bilayer model with intralayer spin-orbit coupling
and contrary Zeeman energy splitting for different layer, which has
been suggested recently to realize 2D time-reversal-invariant
topological superfluid \cite{bbhuang}. Let $H_i$ and $\mathcal{V}_i$
denote single particle Hamiltonian and pairings for ith bilayer,
$H_{i,i+1}$ the coupling between ith and (i+1)th bilayers, the
Hamiltonian of 3D system is \begin{eqnarray}H&=&\sum_i
\left[H_i+\mathcal{V}_i+H_{i,i+1}\right],\label{1'}\\H_i&=&\sum_{{\bf
k_{\bot}}\alpha\beta s}\psi_{{\bf k_{\bot}}i \alpha
s}^\dagger\left\{ \left[\epsilon_{{\bf k_{\bot}}} + \Gamma_{\alpha}
\sigma_z + \lambda_y k_y \sigma_x\right.\right.\nonumber\\
&&\left.\left.-\lambda_x k_x\sigma_y
\right]_{ss'}\delta_{\alpha\beta}-t(1-\delta_{\alpha\beta})\delta_{ss'}\right\}
\psi_{{\bf k_{\bot}}i \beta s'},\nonumber\\
\mathcal{V}_i&=&-\Delta\sum_{{\bf k_{\bot}}\alpha} \psi_{{\bf
k_{\bot}}i\alpha \uparrow}^{\dag}\psi^{\dag}_{-{\bf k_{\bot}}i\alpha
\downarrow}+h.c.,\nonumber\\H_{i,i+1}&=&-t\sum'_{{\bf k_{\bot}}s}
\psi_{{\bf k_{\bot}}i\alpha s}^\dagger \psi_{{\bf k_{\bot}}i+1\beta
s}+h.c.\nonumber\end{eqnarray} where $\psi_{{\bf k_{\bot}}i\alpha
s}^\dagger$ is the creation operator for the fermion particle with
momentum ${\bf k_{\bot}}$, spin $s = \uparrow, \downarrow$ and
sublayer index $\alpha = 1, 2$ in a bilayer. $\epsilon_{{\bf
k_{\bot}}}={\bf k^2_{\bot}}/2m-\mu$ is in-plane kinetic energy
measured from chemical potential $\mu$. The prime over the summation in
$H_{i,i+1}$ means that the sublayer indexes are limited to $\alpha=2$ and $\beta=1$, i.e. we only
consider the nearest neighborhood tunneling between the bilayers.
$t$ is the tunneling between different sublayers. We assume the distance between any two
nearest layer to be $a/2$, so the length for unit cell in $z$
direction is $a$. $\lambda_x$ and $\lambda_y$ are anisotropic
intralayer spin-orbit coupling. Our main investigations are on the isotropic case $\lambda_x=\lambda_y=\lambda$.
We will briefly discuss the effects of the anisotropy. $\Gamma_{\alpha}$
are sublayer-dependent effective Zeeman energy splittings
$\Gamma_1=-\Gamma_2=\Gamma$. Such Zeeman field configuration can be relevant to Van der Waals layered material MnBi$_2$Te$_4$ \cite{zhongdongqin, lijiaheng, mikhail}, in which the intralayer exchange coupling is ferromagnetic, giving 2D ferromagnetism in their septuple layer;
while the interlayer exchange coupling is antiferromagnetic, forming 3D A-type antiferromagnetism
in their bulk. On the other hand this required Zeeman field may also be realized in
ultracold atoms using spin-dependent optical lattice \cite{bbhuang, detush}. Under the
Nambu basis $\Phi_{{\bf k}} = (\psi_{{\bf k}}, \psi_{-{\bf
k}}^{\dag})^T$ with $\psi_{{\bf k}} = (\psi_{{\bf k}1\uparrow},
\psi_{{\bf k}1\downarrow}, \psi_{{\bf k}2\uparrow}, \psi_{{\bf
k}2\downarrow})$ and ${\bf k}=({\bf k_{\bot}}, k_z)$, the Hamiltonian $H$ can be arranged into the
following Bogoliubov-de Gennes (BdG) equation $H = {1 \over
2}\sum_{{\bf k}} \Phi_{{\bf k}}^\dagger H({\bf k}) \Phi_{{\bf k}}$
with
\begin{eqnarray}
H({\bf k}) =
\begin{pmatrix}
\mathcal{H}_0({\bf k}) &  \hat{\Delta} \\
\hat{\Delta}^\dagger & -\mathcal{H}_0^{\ast}(-{\bf k})
\end{pmatrix},\label{1}
\end{eqnarray}
where $\mathcal{H}_0({\bf k})=\epsilon_{{\bf k_{\bot}}}s_0 \otimes
\sigma_0+\Gamma s_z\otimes \sigma_z-2t \cos{(k_z a/2)}
s_x\otimes\sigma_x+\lambda_y k_y s_0\otimes\sigma_x- \lambda_x k_x
s_0\otimes \sigma_y$, $\hat{\Delta} = -i\Delta s_0\otimes\sigma_y$.
$s_{0,x,y,z}$, $\sigma_{0,x,y,z}$ are two sets of Pauli matrices
acting on the layer and spin spaces. We will also introduce
$\tau_{0,x,y,z}$ as Pauli matrix acting on the particle-hole space.
A feature of this Hamiltonian is that it is free in $xy$ plane and
constrained by a lattice in $z$ direction.

The Hamiltonian (\ref{1}) has PHS $\Sigma H^{\ast}({\bf
k})\Sigma^{-1}=-H(-{\bf k})$ with $\Sigma=\tau_x\otimes
s_0\otimes\sigma_0$ and TRS $\mathcal{T}H^{\ast}({\bf
k})\mathcal{T}^{-1}=H(-{\bf k})$ with $\mathcal{T}=i\tau_0\otimes
s_x\otimes\sigma_y$. Thus our model belongs to 3D DIII class. The
combination of TRS and PHS gives rise to a chiral symmetry
$C=i\mathcal{T}\Sigma$ with $C H({\bf k})C^{-1}=-H({\bf k})$. In
addition, the model (\ref{1}) also has spatial inversion symmetry $I
H({\bf k})I^{-1}=H(-{\bf k})$ with $I=\tau_z\otimes
s_0\otimes\sigma_z$ and mirror symmetries $\mathcal{M}_x
H(k_x,k_y,k_z)\mathcal{M}^{-1}_x=H(-k_x,k_y,k_z)$, $\mathcal{M}_y
H(k_x,k_y,k_z)\mathcal{M}^{-1}_y=H(k_x,-k_y,k_z)$, $\mathcal{M}_z
H(k_x,k_y,k_z)\mathcal{M}^{-1}_z=H(k_x,k_y,-k_z)$ with
$\mathcal{M}_x=I\mathcal{T}$, $\mathcal{M}_y=-i\mathcal{T}$ and
$\mathcal{M}_z=\tau_0\otimes s_0\otimes\sigma_0$. These mirror
symmetries are very vital to topological stabilities of MFA in the
model (\ref{1}), as demonstrated below.

\begin{figure}
\includegraphics[width=8.0cm,height=12.0cm]{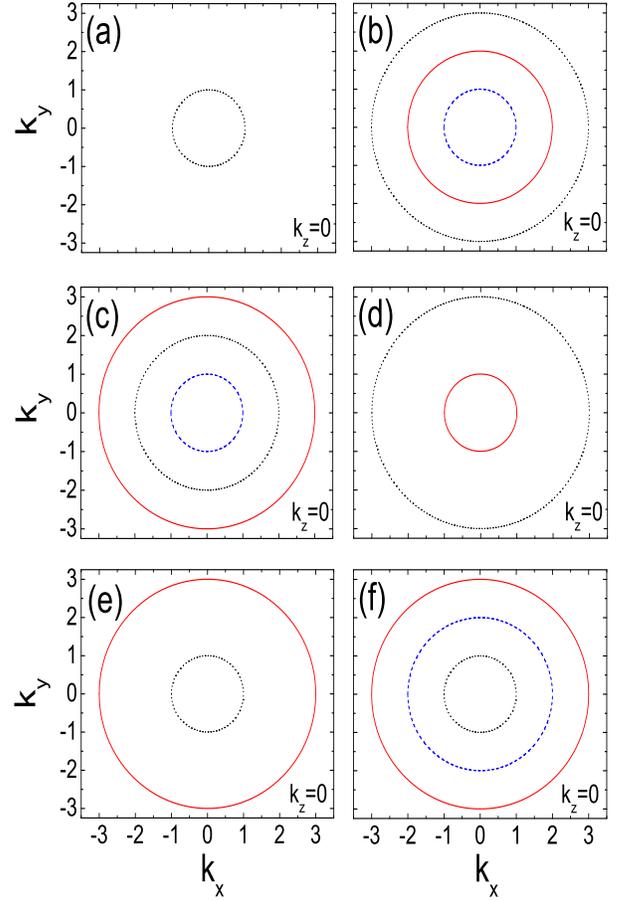}
\caption{Possible phases in the model (\ref{1}). The schematic plots
for two equations in (\ref{3}) at $k_z=0$ plane. The radius of
black circle (dotted) is $2t/\lambda$, while the radii of red (solid) and blue (dashed)
circles are $\sqrt{W_c^{+}}$, $\sqrt{W_c^{-}}$ if they can be
defined. In 3D the red (blue) circle corresponds to an
open cylindrical surface along the $k_z$ axis, while the black circle
gradually decreases to zero when $k_z$ reaches the boundary of
Brillouin zone to form an irregular spherical surface. Thus as long as one of blue and red circles is into the black one, the cylindrical surface(s) will cross irregular spherical surface to form gapless nodal lines of the model (\ref{1}). From this criterion (a), (e), (f) are gapped phases, (b) is gapless phase with
four nodal lines  and (c), (d) are
gapless phases with two nodal lines.  In the maintext, (a) [(e) and (f)] is (are) denoted by
gapped phase I (III), while (c) and (d) [(b)] are (is) denoted by gapless
phase II (IV).}
\end{figure}

\section{Possible phases and bulk topology} The structure of phase
diagram can be obtained from analyzing the determinant of $H({\bf
k})$, which can be calculated with the help of the chiral symmetry.
By finding a unitary matrix $U=(\tau_x\otimes s_x\otimes
\sigma_y+\tau_z\otimes s_0\otimes \sigma_0)/\sqrt{2}$ to diagonalize
$C$ that $ UCU^\dagger =\tau_z\otimes s_0\otimes\sigma_0$, we can
transform the Hamiltonian into an off-diagonal form
\begin{equation}
U H({\bf k}) U^\dagger =
\begin{pmatrix}
0  & Q \\
Q^\dagger & 0
\end{pmatrix}\label{2},
\end{equation}
with $Q=\mathcal{H}_0({\bf
k})\cdot(s_x\otimes\sigma_y)-\hat{\Delta}$. Thus $\text{Det}(H({\bf
k})) = \text{Det}(Q) \cdot \text{Det}(Q^\dagger)\geq 0$ and the gap
closing condition is $\text{Det}(Q)= 0$. Following this method, we
find for our model gap closing signifies that the equations
\begin{eqnarray}
\epsilon_{{\bf k_{\bot}}}^2+\Delta^2=\Gamma^2, \lambda^2{\bf k_{\bot}}^2=4t^2\cos^2{\frac{k_za}{2}}
\label{3}
\end{eqnarray}
have solutions for ${\bf k_{\bot}}$ and $k_z$. With the second
equation in (\ref{3}) defining an irregular spherical surface, the
first equation defines possible cylindrical surfaces depending on
the values $W_c^{\pm}=2m(\mu\pm\sqrt{\Gamma^2-\Delta^2})$. Thus the
solutions of (\ref{3}) correspond to the intersections of spherical
surface and cylindrical surfaces and define some gapless nodal
lines. Since the number of cylindrical surfaces is at most equal to
2, the maximal number of nodal lines is 4. The relation between all
possible phases and parameters $2t/\lambda$, $W_c^{\pm}$ can be
found in Fig.1. Here we notice that the sign of $\mu$ affects the
phases available. For $\mu<0$ the maximal number of phases is 3, but
for $\mu>0$ the maximal number of phases is 4.

Below we discuss the topological properties of all phases from the
3D bulk viewpoint. For the gapless phases, there are pairwise nodal
lines. The featured structures of nodal lines can be visualized from
the circles at $k_z=0$ plane in Fig.1. To characterize their
stabilities, we calculate the winding number \cite{wangzd1, beri,
sato} $N_1=\frac{1}{2\pi}\oint_{S^1} d{\bf k}\cdot
\partial_{{\bf k}}\text{Im}[\ln\text{Det}(Q)]$, where the contour
$S^1$ encircles the one of nodal lines from its transverse
direction. A nonzero $N_1$ signifies topological stability. In our
model $N_1= \pm 1$. The topological superfluid with nodal lines can
be regarded as the supefluid counterpart of nodal line semimetal
widely studied in the solid state materials Ca$_3$P$_2$ \cite{nl1},
SrIrO$_3$ \cite{nl2}, PbTaSe$_2$ \cite{nl3} and stable 3D carbon
allotrope \cite{nl4, nl5}. While for gapped phase, we could consider
a strong topological invariant \cite{Schnyder08} $N_2=\int\frac{ d^3
{\bf k}}{24 \pi^2}\epsilon^{ijk} \text{Tr}[(Q^{-1}\partial_i
Q)(Q^{-1}\partial_j Q)(Q^{-1}\partial_k Q)] $ with
$i,j,k=k_x,k_y,k_z$ defined in the whole momentum space, but $N_2=0$
due to trivial mirror symmetry $\mathcal{M}_z$ in our model, which
leads to $Q(-k_z)=Q(k_z)$ \cite{zhangfan}. $N_2=0$ means our model
cannot realize a strong 3D DIII class topological superfluid, but as illustrated
in the next section, the gapped phase III owns anomalous surface
states in $xz$ and $yz$ side surfaces, thus it remains topological.

\section{Majorana Fermi arc and topological protection} In this
section surface states on the $yz$ side surface are discussed. We
consider a slab with periodic boundary condition in $y$, $z$
directions and a finite thickness $L$ along $x$ direction. Expanding
the wavefunction $\psi(x) =\sum_{n=1}^{\infty} c_n \sin{(n \pi
x/L)}$ \cite{dong, huang}, the calculated results for $k_y=0$ are
presented in Fig.2. There exists MFA with four fold degeneracy in two gapless phases and
gapped phase III.

\begin{figure}
\includegraphics[width=8.0cm,height=7.0cm]{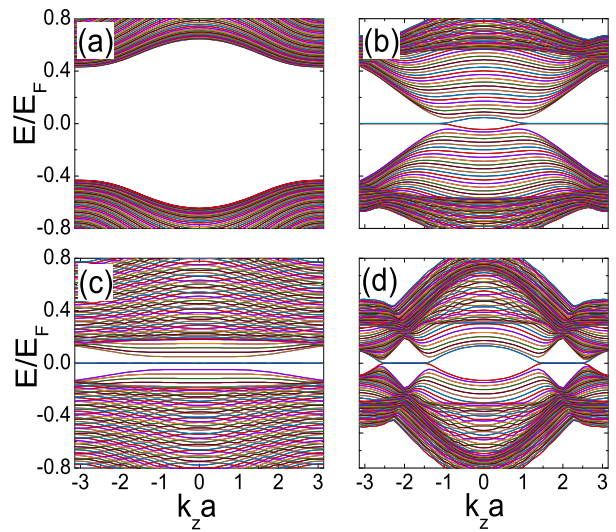}
\caption{Surface states on $yz$ side surfaces at $k_y=0$ for four
phases in our model (\ref{1}). (a)-(d) correspond to phase I-IV,
respectively. In (a) $\Gamma=0.8E_F$, $\Delta=1.23E_F$,
$\mu=-0.07E_F$, (b)  $\Gamma=1.3E_F$, $\Delta=0.63E_F$,
$\mu=-0.11E_F$, (c)  $\Gamma=1.8E_F$, $\Delta=0.22E_F$,
$\mu=-0.43E_F$, (d)  $\Gamma=0.41E_F$, $\Delta=0.23E_F$,
$\mu=0.42E_F$, with $t=0.55E_F$, $\lambda k_F=1.0E_F$. For $E_F$ and $k_F$, see Fig.3.}
\end{figure}

To correctly illustrate the observed MFA, we introduce two
different topological invariants. Firstly we regard 3D system as a
set of 2D subsystems, parameterized by the momentum component $k_z$.
Generally the symmetries satisfied by a 3D system must not remain
symmetries of 2D subsystem, especially for TRS and PHS, since they
correlate the momentum $\textbf{k}$ with $-\textbf{k}$ non-locally.
However in our model due to trivial MS $\mathcal{M}_z$
\cite{furusaki}, TRS $\mathcal{T} H({\bf
k_{\bot}},k_z)\mathcal{T}^{-1}=H(-{\bf k_{\bot}},k_z)$ and PHS
$\Sigma H({\bf k_{\bot}},k_z)\Sigma^{-1}=-H(-{\bf k_{\bot}},k_z)$
always exist for 2D subsystems. Thus every $k_z$ fixed plane belongs
to 2D DIII class and its topological properties can be distinguished
by the $Z_2$ number \cite{qixiaoliang}
\begin{eqnarray}
\nu_2(k_z)=\prod_j[\text{sgn}
(\delta_j)]^{p_j}, \label{4}
\end{eqnarray}
where $p_j$ is the number of time-reversal-invariant points enclosed
by the $j$-th Fermi surface, $\delta_{j,\bold{k}} = \langle j
\bold{k}|(is_x\otimes\sigma_y)\cdot\hat{\Delta}^{\dag}|j \bold{k}
\rangle$ and sgn$(\delta_j)$ is the sign of pairing gap on the Fermi
surface with $|j\bold{k}\rangle$ being the eigenstates of
$\mathcal{H}_0(\bold{k})$.

Additionally we consider the mirror symmetry $\mathcal{M}_y$. For
$k_y=0$, the Hamiltonian commutes with $\mathcal{M}_y$ and can be
partitioned into two independent mirror subspaces
$M_{\pm}(k_x,k_z)=-\epsilon_{k_x}\sigma_z\otimes \sigma_0-\Gamma
\sigma_z\otimes \sigma_z-(\lambda k_x\pm 2t
\cos{\frac{k_za}{2}})\sigma_z\otimes \sigma_y+\Delta
\sigma_y\otimes\sigma_y$ under the basis of diagonalizing
$\mathcal{M}_y$ \cite{ms1, ms2, ms3, ms4, ms5}. Since
$\mathcal{M}_y$ anticommutes with TRS and PHS, every mirror subspace
belongs to AIII class with the common chiral operator
$S=\sigma_x\otimes\sigma_0$. Thus every mirror subspace can be
further written into the nondiagonal form
\begin{eqnarray}
M_{\pm}(k_x,k_z)\sim \begin{pmatrix}
0  & Q_{\pm} \\
Q_{\pm}^\dagger & 0
\end{pmatrix} \label{6}
\end{eqnarray}
with $Q_{\pm}=\epsilon_{k_x}\sigma_0-\Gamma \sigma_z-(i\Delta+\lambda k_x \pm 2 t \cos{\frac{k_za}{2}})\sigma_y$. In the mirror subspaces, every 1D subsystem can be classified by the winding number
\begin{eqnarray}
N_{\pm}(k_z)=\frac{1}{2\pi}\int_{-\infty}^{\infty}dk_x \partial_{k_x}\text{Im}[\ln\text{Det} Q_{\pm}]. \label{7}
\end{eqnarray}
Since two mirror subspaces are correlated with each other by TRS and PHS, the winding numbers $N_{\pm}(k_z)$ are not independent and we can prove exactly $N_{+}(k_z)=N_-(k_z)$.

We have numerically calculated $\nu_{2}(k_z)$, $N_{\pm}(k_z)$ and
find that two topological invariants are nonzero as long as
$\Gamma^2>
\left[2t^2\cos^2{\frac{k_za}{2}}/(m\lambda^2)-\mu\right]^2
+\Delta^2$. In
gapped phase I (III), two invariants are trivial (nontrivial) for all $k_z$ fixed subsystems; In
gapless phase II, only subsystems with
$k_z\in(-\pi/a,-k_z^+)\cup(k_z^+,\pi/a)$ or $k_z\in(-k_z^-, k_z^-)$ are nontrivial, where
$\cos{(k_z^{\pm}a/2)}=\sqrt{\lambda^2W_c^{\pm}/(4t^2)}$, depending on that the gapless phase II corresponds to Fig.1(b) or Fig.1(d); While in gapless IV only subsystems with
$k_z\in(-k_z^-,-k_z^+)\cup(k_z^+,k_z^-)$ are nontrivial. We also find that the range in which two topological invariants are nonzero is exactly the same as that MFA exist, in other words
MFA in the  model (\ref{1}) is protected by $\nu_{2}(k_z)$,
$N_{\pm}(k_z)$. No matter which topological invariant protect the
MFA in Fig.2, it is different from MFA in the time-reversal broken
Weyl superfluid protected by nonzero Chern number. This important
observation means that MFA in our model represents a new type of
topological state not explored previously. The same MFA also exists
in $xz$ surface and can be analyzed similarly if we consider the mirror symmetry $\mathcal{M}_x$.

\section{phase diagram with self-consistent pairing gap}

Within BCS mean-field theory, let $-U$ ($U>0$) denote the effective attraction strength in each layer. Then the pairing gaps
$\Delta_{\alpha} = U\sum_{{\bf k}}\langle \psi_{-{\bf k}\alpha\downarrow}
\psi_{{\bf k}\alpha\uparrow} \rangle$ and the free energy at zero temperature
$F = {1\over 2}\sum_{{\bf k}} \left[\text{Tr}\,
\mathcal{H}_0({\bf k})-\sum_{\eta=1}^4 E_{\eta{\bf k}}\right] +
(|\Delta_1|^2+|\Delta_2|^2)/U+ \mu N$. Here $E_{\eta{\bf k}}$ are four positive eigenvalues of BdG Hamiltonian $H({\bf k})$ and $N$ is the particle number.
The integral for ${\bf k}_\bot$ is divergent and we choose to use the relation
$1/U=\sum_{{\bf k_{\bot}}}({\bf k^2_{\bot}}/m+\varepsilon_b)^{-1}$ with
$\varepsilon_b$ being the binding energy in 2D free space
\cite{randeria} to regularize $F$. The physical properties of the ground
state can be obtained by minimizing $F$. Notice that the conclusions of nodal line topological superfluid and multiply protected MFA in model (\ref{1}) is highly depend on whether the order parameters $\Delta_{\alpha}$ of two layers are equivalent.
The same magnitudes for $\Delta_{\alpha}$ can be directly anticipated from the bilayer's symmetry. In order to check whether $\Delta_{\alpha}$ have the same phases, we consider the ansatz
$\Delta_1=\Delta$, $\Delta_2=\Delta e^{i\phi}$ to investigate the behavior of free energy $F$
as the function of the relative phase $\phi$ by self-consistently solving order parameter $\Delta$ and chemical potential $\mu$.
We find that $F$ always obtains minimum at $\phi=0$. Fig.3(a) shows the landscapes of the free energy for two chosen parameter sets. When
$\Delta_1=\Delta_2$, TRS $\mathcal{T}$ and chiral symmetry
$C$ are recovered and our conclusions about bulk
topology and MFA are valid. The phase diagram from the
self-consistent calculation is shown in Fig.3(b).

\begin{figure}
\includegraphics[width=8.0cm,height=4.0cm]{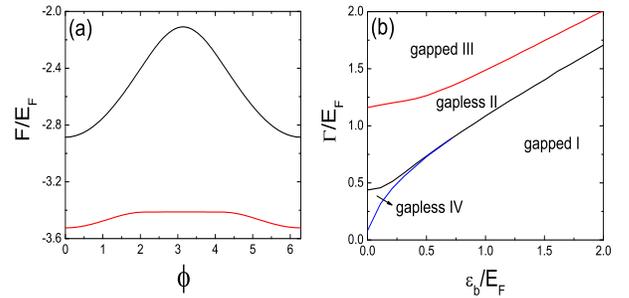}
\caption{The landscapes of free energy $F$ as the function of the relative phase $\phi$ of two order parameters $\Delta_{\alpha}$ (a) and phase diagram from the self-consistent calculation (b). In (a) $\varepsilon_b=1.0E_F$, $\Gamma=0.8E_F$ for the black (upper) line and  $\Gamma=1.3E_F$ for the red (lower) line. The step structure in the middle of red (lower) line corresponds to the normal state with $\Delta_{\alpha}=0$. The
other parameters are $t=0.55E_F$, $\lambda k_F=1.0E_F$. We define the Fermi momentum $k_\text{F} = \sqrt{\pi n}$
from particle number density $n$ for every bilayer and Fermi energy
$E_F = k^2_F/2m$ as the unit of inverse length and energy. For the definitions of different phases, see Fig.1.}
\end{figure}

\section{Discussions}

We want to emphasize the following comments.
Let us note that the model (\ref{1}) also have the chiral symmetry
$C$. Thus at $k_y=0$ we can define similar topological invariant
(\ref{7}) by substituting $Q$ for $Q_{\pm}$. The numerical results
suggest that this winding number is often zero and cannot protect
the MFA observed in Fig.2. (This winding number in essence is the
same as $N_1$, but its integral path cannot embrace the nodal
lines.) However this chiral symmetry can protect some other
interesting surface states. For example when the projections of
nodal lines on certain surfaces do not overlap completely, there are
zero-energy Andreev bound states on this surface located within the
projected nodal rings \cite{ryu}. Here we are not interested in
them.

In our Hamiltonian, Rashba spin-orbit coupling has been assumed,
which ensures flat nodal lines and the same MFA in $yz$ and $xz$
surfaces. However the existences of gapless nodal lines and MFA are
also robust for the anisotropic spin-orbit coupling. When we change
$\lambda_y$ ($\lambda_x$) slightly, every flat nodal line will
develop into a spatial curve and have a finite extension along $k_z$
axis, while the MFA on $yz$ ($xz$) surfaces will not change when other
parameters are fixed. But for a large anisotropy, the number of
nodal lines and MFA can change once the red or blue circle
intersects with the black ellipse at $k_z=0$ in Fig.1. As an
example, let us decrease $\lambda_y$ from Fig.1(e) to induce the
crossings. As a result gapped phase III transits into gapless phase
II and the MFA in $xz$ surface has the similar structure with
Fig.2(b), with the MFA in $yz$ surface invariant. Since a nonzero 2D
topological invariant $\nu_2(k_z)$ means the simultaneous presence
of edge states in $yz$ and $xz$ surfaces, thus generally for
$\lambda_x\ne\lambda_y$ the MFA is only protected by $N_{\pm}(k_z)$. Moreover the above example also demonstrates
that for $\lambda_x\ne\lambda_y$ the different phases can own the
same MFA in $yz$ or $xz$ surface.

One of the motivations considering sublayer-dependent effective Zeeman energy splittings
$\Gamma_1=-\Gamma_2=\Gamma$ comes from the recent discovery of layer antiferromagnetic topological insulator MnBi$_2$Te$_4$ \cite{zhongdongqin, lijiaheng, mikhail}. In the low energy effective model of MnBi$_2$Te$_4$, there are $4$ bands for every septuple layer, thus for the antiferromagnetic bilayer there are $8$ bands totally; While we only consider a two-band case for the single layer, which is usually used to describe semiconductors or Fermi gases with spin-orbit coupling and is also the starting point for studying D class topological superfluid in 2D \cite{dc1, dc2, dc3}. Additionally from the 3D perspective the single-particle model of MnBi$_2$Te$_4$ has been topological, but it is not true for our model. Although the model (\ref{1}) is more simple, it has shown lots of exotic quantum phenomena, as demonstrated before. Thus to investigate superconducting effects in the realistic antiferromagnetic topological insulator should be deserved. The superconductivities in time-reversal-invariant topological insulator, Dirac and Weyl semimetal have been widely studied and bring interesting results \cite{ti1, ti2, ti3, dw}.

Experimentally, four kinds of
phases in our model can be solely identified by radio-frequency
spectroscopy in the cold atom gases and angle-resolved photoemission spectroscopy in condensed matter materials \cite{zhengkun, zhangjing, jiangkaijun, arpes1, arpes2, arpes3}. On one hand
gapless and gapped phases can be distinguished by nodal lines in the
gapless phases, meantime different number of nodal lines can also
discriminate different gapless phases. On the other hand in contrast
to the gapped phase I, the existence of MFA in gapped phase III
enhances the low-energy spectral function and becomes a smoking gun
for identifying the different gapped phases.

\section{Conclusion}

To conclude we theoretically study a
time-reversal-invariant three-dimensional superfluid model by
stacking a series of bilayers with intralayer spin-orbit coupling
and contrary Zeeman energy splitting for different layer. This model
shows two kinds of topologically nontrivial phases: gapless phases
with nodal lines in pairs protected by chiral symmetry and a gapped
time-reversal-invariant weak topological superfluid phase, both of
which support time-reversal-invariant Majorana Fermi arc on the $yz$
and $xz$ side surface, which is topologically protected by the
winding numbers in mirror subspaces and $Z_2$ numbers of two-dimensional DIII class topological
superfluid, different from Majorana Fermi arc in the time-reversal
broken Weyl superfluid protected by nonzero Chern number. The exotic
Majorana Fermi arc predicted here represents a new type of
topological state and provides fertile grounds for exploring exotic
quantum matters.

\textit{Acknowledgements.} We would like to thank Ming Gong for discussions. This
work is supported by NSFC under Grant No. 11547047 and No. 11704324.

\end{document}